\documentstyle[12pt,a4]{article}
\begin{document}
\baselineskip=17pt plus 0.2pt minus 0.1pt
%\baselineskip=19pt plus 0.2pt minus 0.1pt

%%%%%%%%%% Private macros %%%%%%%%%%%%
\makeatletter
\def\CN{{\cal N}}
\def\d{{\rm d}}
\def\Tr{\mathop{\rm Tr}}
\def\tr{\mathop{\rm tr}}
\def\cc{{g_{\rm YM}^2}}
\def\calO{{\cal O}}
\def\calL{{\cal L}}
\def\calD{{\cal D}}
\def\del{{\partial}}
\def\half{{1 \over 2}}
\def\nn{{\nonumber}}
\newcommand{\cd}{\!\cdot\!}
\newcommand{\equal}{\!\!\!&=&\!\!\!}
\newcommand{\dlt}{\delta_{\rm SCT}^{\rm C}}
\newcommand{\Dlt}{\delta_{\rm SCT}^{\rm Q}}
\newcommand{\bra}[1]{\left\langle #1\right|}
\newcommand{\ket}[1]{\left| #1\right\rangle}
\newcommand{\VEV}[1]{\left\langle #1\right\rangle}
\newcommand{\braket}[2]{\VEV{#1 | #2}}
\newcommand{\diag}{\mathop{\rm diag}}
\newcommand{\bb}{\!\!\!}
\newcommand{\Deta}{\Delta^{\eta}}
\newcommand{\dltot}{\delta_{\rm SCT}}
\newcommand{\Goneloop}{\Gamma_{1\hbox{\scriptsize -loop}}}
\newcommand{\bea}{\begin{eqnarray}}
\newcommand{\eea}{\end{eqnarray}}
\newcommand{\be}{\begin{equation}}
\newcommand{\ee}{\end{equation}}
\newcommand{\dBRST}{\delta_{\rm BRST}}
\newcommand{\barC}{\bar{C}}
\newcommand{\ds}{\displaystyle}
%%%%%%%%%%%%%%% dddot and ddddot %%%%%%%%%%%%%%%%%%%%
\newdimen\ex@
\ex@.2326ex
\def\dddot#1{\raise\ex@\hbox{${\mathop{#1}\limits^{
             \vbox to-1.4\ex@{\kern-\tw@\ex@\hbox{\rm...}\vss}}}$}}
\def\ddddot#1{\raise\ex@\hbox{${\mathop{#1}\limits^{
             \vbox to-1.4\ex@{\kern-\tw@\ex@\hbox{\rm....}\vss}}}$}}
%%%%%%%%%%%%%%%%%%%%%%%%%%%%%%%%%%%%%%%%%%%%%%%%%%%%%%

\makeatother

%%%%%%%%%% End of private macros %%%%%%%%%%%%

\begin{titlepage}

\title{
\hfill\parbox{4cm}
{\normalsize KUNS-1568\\{\tt hep-th/9904042}}\\
\vspace{1cm}
Conformal Symmetry and A New Gauge\\
in the Matrix Model
}
\author{
Hiroyuki {\sc Hata}\thanks{{\tt hata@gauge.scphys.kyoto-u.ac.jp}
}
{} and
Sanefumi {\sc Moriyama}
\thanks{{\tt moriyama@gauge.scphys.kyoto-u.ac.jp}
}
\\[7pt]
{\it Department of Physics, Kyoto University, Kyoto 606-8502, Japan}
}
\date{\normalsize April, 1999}
\maketitle
\thispagestyle{empty}

\begin{abstract}
\normalsize
We generalize the background gauge in the Matrix model to propose a
new gauge which is useful for discussing the conformal symmetry.
In this gauge, the special conformal transformation (SCT) as the
isometry of the near-horizon geometry of the D-particle solution is
directly reproduced with the correct coefficient as the quantum
correction to the SCT in the Matrix model.
We also present a general argument for the relation between the gauge
choice and the field redefinition in the Matrix model.
\end{abstract}

\end{titlepage}

\section{Introduction}
It has been established that at low energies D-branes are described
effectively by the Born-Infeld action, whose lowest order with respect
to the number of derivatives is the Super-Yang-Mills theory (SYM)
\cite{Witten}.
When M-theory is compactified in the infinite momentum
frame, only the degrees of freedom of D-particles remain.
It was therefore conjectured that the system is fundamentally
described by $d=1$ SYM, i.e., the Matrix model \cite{BFSS,Susskind}.
The conjecture of the Matrix model has been intensively investigated
in many systems and compared to supergravity.
Especially, D-particle scatterings
with multi-bodies have been checked up to two loops \cite{DKPS,BBPT,OY},
though disagreement for higher loops in a more complicated system is
suspected \cite{DEG}. Therefore, we have to show the full agreement
between the Matrix model and supergravity to confirm the validity of
the Matrix model as a non-perturbative definition of M-theory.
A possible way to show the full agreement is to rely on symmetries.
For example, the supersymmetries put restrictions on the form of the
Matrix model effective action in the first few orders of the
derivative expansion \cite{SSP,SC,O}.

Conformal symmetry imposes another restriction on the effective action
of the SYM, and in particular, the Matrix model.
In fact, it is shown in \cite{Malda} that the Born-Infeld action
with the background of the near-horizon geometry of the D3-brane
solution (that is, the Anti-de-Sitter or AdS space) can be determined
exactly by the isometry of the AdS space.
The special conformal transformation (SCT) of the isometry of the AdS
space differs from the canonical one in SYM by an extra term which
vanishes on the boundary. However, interestingly, this extra
term can be derived from SYM as a quantum correction
\cite{JKY3}.
In this way, the problem of showing the full agreement between the
Born-Infeld action with the AdS background and the effective action of
SYM is reduced to showing that the quantum modified symmetry
of SYM reproduces exactly the isometry of the AdS space.

Although the maximally SYM in $d\ne 4$ is not a conformal field theory
and the near-horizon geometry of the D$p$-brane with $p\ne 3$
is not of the AdS type, one can generalize the above arguments to
D$p$-branes by varying also the coupling constant under the dilatation
and the SCT in both the SYM and the D$p$-brane geometry \cite{JY}.
This time, though the isometry of the near-horizon geometry determines
the Born-Infeld action, the SCT derived from the SYM appears
to differ from that of the isometry by a numerical factor.
The SCT Ward-Takahashi identity still holds only when one keeps the
terms proportional to the derivative of the coupling constant in the
effective action, because they are also relevant after the
transformation \cite{JKY0,HM}.
It is also shown that the two SCTs (i.e., the isometry and the quantum
modified one in SYM) are related by a field redefinition.
Though consistent, the notorious numerical factor prevents us
from determining the Born-Infeld action directly.
Hence, it would be difficult to show the full agreement between
the Matrix model and the Born-Infeld theory from the symmetry.
Besides, it is unclear why the field redefinition is necessary in
spite of the fact that in other works \cite{DKPS,BBPT,OY} studying the
relationship between the Matrix model and the supergravity they always
agree without any field redefinitions. 

In this paper, we present a new gauge in SYM, which is a natural
extension of the background gauge adopted in all the previous works
\cite{JY,JKY0,HM}. This new gauge has a marvelous property that
the SCT as the isometry of the near-horizon geometry of the D$p$-brane
solution is correctly reproduced without field redefinitions as the
quantum modified SCT of SYM, unlike in the case of the conventional
background gauge.

Then, the result in our new gauge raises a question: what is
the meaning of the gauge choice in the Matrix model? Therefore,
we give a general argument to identify the change of gauge-fixing in
the Matrix model as the redefinition of the D-brane coordinates.
As a concrete example of this argument, we carry out the calculation
in the $R_\xi$ gauge. We find among other things that the 
agreement between the Matrix model and the supergravity without any
field redefinitions \cite{DKPS,BBPT,OY} is merely a special nature of
the background gauge.

The organization of the rest of this paper is as follows.
In section 2, we summarize the quantum conformal symmetry in SYM.
We derive the quantum modified SCT and explain the disagreement of the
numerical factor with that in the isometry.
In section 3, we present our new gauge and analyze the quantum SCT in
this gauge.
In section 4, we give a general argument on the relation of
changing the gauge in SYM to the field redefinition in the effective
action, and in section 5 we analyze the $R_\xi$ gauge as an example
of the general formalism given in section 4.
In the final section, we summarize the paper and discuss further
directions. In the appendices, we present some technical details used
in the text.

\section{Quantum conformal symmetry in SYM}
First, let us summarize the quantum conformal symmetry in SYM
\cite{JY,JKY3,JKY0,HM} in detail, because the methods needed
afterwards are essentially the same.
It is well known that the action of $d=4$ $\CN=4$ SYM has
conformal symmetry.
This is also the case for the SYM with sixteen supersymmetries in
other dimensions if we assign the coupling constant a conformal
dimension and vary it under the transformation.
We will here consider SYM in the Euclidean formulation:
\begin{equation}
S_{\rm SYM}=\int\!\d^{p+1}\!x\,\frac{1}{\cc}
\tr\left(\frac{1}{4}F_{\mu\nu}^2
+\half(D_\mu X_m)^2-\frac{1}{4}[X_m,X_n]^2+
\mbox{fermionic part}
\right).
\label{SYM}
\end{equation}
The action (\ref{SYM}) is invariant under both the dilatation
and SCT. In particular, the transformation law of the bosonic
variables under SCT reads
\begin{eqnarray}
&&\dlt x_\mu=2\epsilon\cd xx_\mu-\epsilon_\mu x^2,\quad
\dlt A_\mu=-2\epsilon\cd xA_\mu
-2(x\cd A\epsilon_\mu-\epsilon\cd Ax_\mu),\nn\\
&&\dlt X_m=-2\epsilon\cd xX_m,\quad
\dlt\cc=2(p-3)\epsilon\cd x\cc,\label{SCT}
\end{eqnarray}
where the superscript C in $\dlt$ is for distinguishing the present
classical transformation from the quantum one to be given later.

In order to quantize the system, we have to add the gauge-fixing and
the corresponding ghost terms to the original action (\ref{SYM}):
\bea
S_{\rm gf+gh}=\int\!\d^{p+1}\!x\,i\delta_{\rm BRST}\tr(\bar{C}G).
\label{gf+gh}
\eea
In the original work of \cite{JKY0}, they adopted the famous
background gauge
\begin{equation}
G=-\del_\mu A_\mu+i[B_m,Y_m]+\half\cc b,
\label{BGgauge}
\end{equation}
where $Y_m$ is the fluctuation of the scalars $X_m$ from the diagonal
background $B_m$; $X_m=B_m+Y_m$, and $b$ is the auxiliary field for
the off-shell closure of the BRST algebra.
The BRST transformation of the fields are
\bea
&&\dBRST A_\mu= D_\mu C\equiv \del_\mu C -i[A_\mu,C],\quad
\dBRST X_m=-i[X_m,C],\nn\\
&&\dBRST C=i C^2,\quad
\dBRST\barC=i b,\quad
\dBRST b=0.
\label{dBRST}
\eea
We assign the SCT of the unphysical fields so that $\dlt$ and the BRST
transformation $\dBRST$ are commutative, $[\dlt,\dBRST]=0$:
\bea
\dlt C=0,\quad
\dlt\barC=-2(p-1)\epsilon\cd x\,\barC,\quad
\dlt b=-2(p-1)\epsilon\cd x\,b.
\label{unphysSCT}
\eea

While $S_{\rm gf+gh}$ (\ref{gf+gh}) with $G$ of (\ref{BGgauge}) is
dilatation invariant, it is not invariant under SCT (\ref{SCT}):
\bea
\dlt S_{\rm gf+gh}=i\dBRST\lambda[A,\barC],
\label{dltS}
\eea
where $\lambda[A,\barC]$ is given by
\begin{eqnarray}
\lambda=\frac{p-1}{2}\cd(-4)\int\!\d^{p+1}\!x
\tr\left(\bar{C}(x)A(x)\cd\epsilon\right).
\label{lambda}
\end{eqnarray}
In general, if a symmetry of the classical action is violated by the
gauge-fixing and the ghost terms $S_{\rm gf+gh}$ and the violation is
given as the BRST-exact form $i\dBRST \lambda$, the symmetry can be
restored by adding to the original transformation for a generic field
$\phi$ the BRST transformation $-i\lambda\dBRST\phi$ with this (field
dependent) transformation parameter $\lambda$.
In fact, the change of the path-integral measure under the added BRST
transformation just cancels the violation $i\dBRST \lambda$.
Therefore, in the present case, the effective action $\Gamma[B,\cc]$
of the system satisfies the following SCT Ward-Takahashi identity,
\begin{equation}
\int\!\d^{p+1}\!x\left(\dlt \cc(x)\frac{\delta}{\delta \cc(x)}
+\left(\dlt+\Dlt\right)\!B_{m,i}(x)\,\frac{\delta}{\delta B_{m,i}(x)}
\right)\Gamma[B,\cc]=0,
\label{WI}
\end{equation}
where the extra term $\Dlt B_{m,i}$ for
$B_m=\diag\left(B_{m,i}\right)$ is
\begin{eqnarray}
\Dlt B_{m,i}(x)=2(p-1)\Bigl\langle[C(x),X_m(x)]_{ii}
\int\!\d^{p+1}y\,\tr(\bar{C}(y)A(y)\cd\epsilon)\Bigr\rangle.
\label{extraterm}
\end{eqnarray}
Note that the total SCT for $B_m$ is now given as a sum of the
classical part $\dlt B_{m,i}=-2\epsilon\cdot xB_{m,i}$ and the
quantum correction $\Dlt B_{m,i}$.

Now let us explicitly calculate $\Dlt B_{m,i}$ (\ref{extraterm}). At
the 1-loop order it is given by
\begin{eqnarray}
\Dlt B_{m,i}(x)=2(p-1)\int\!\d^{p+1}\!y\left(
\langle C_{ij}(x)\,\bar{C}_{ji}(y)\rangle
\VEV{Y_{m,ji}(x)A_{\mu,ij}(y)\epsilon_\mu}
-(i\leftrightarrow j)\right),
\label{extraterm_oneloop}
\end{eqnarray}
where the free propagators are
\begin{eqnarray}
\langle C_{ij}(x)\bar{C}_{ji}(y)\rangle\equal
i\bra{x}\Delta_{ij}\ket{y},\nn\\
\VEV{Y_{m,ji}(x)A_{\mu,ij}(y)}\equal
-2i\cc\bra{x}\Delta_{ij}(\del_\nu B_{m,ij})\Delta_{ij}
\left({\cal M}^{-1}\right)_{\nu\mu}\ket{y},\label{freeprop}
\end{eqnarray}
with $\Delta_{ij}\equiv\left(-\del^2+B_{ij}^2\right)^{-1}$,
$B_{m,ij}\equiv B_{m,i}-B_{m,j}$ and
${\cal M}_{\mu\nu}\equiv\delta_{\mu\nu}
-4(\del_\mu B_{\ell,ij})\Delta_{ij}(\del_\nu B_{\ell,ij})\Delta_{ij}$.
Keeping only the lowest order terms in the derivatives,
eq.\ (\ref{extraterm_oneloop}) is reduced to
\be
\Dlt B_{m,i}(x)=\sum_{j}
8(p-1)\cc\epsilon\cd\del B_{m,ij}\Delta_{ij}^3
=\sum_{j}\frac{4(p-1)\Gamma\Bigl((5-p)/2\Bigr)\cc}
{(4\pi)^{(p+1)/2}B_{ij}^{5-p}}\epsilon\cd\del B_{m,ij}.
\label{Dlt_B_final}
\ee
Restricting ourselves to the source-probe configuration with $N$
D$p$-branes as the source at the origin and the probe at $B_m$;
$B_m=\diag(0,\cdots,0,B_m)$,
we obtain the final form of the quantum modified SCT
$\dltot\equiv\dlt+\Dlt$ for $x_\mu$, $\cc$ and $U_m\equiv 2\pi B_m$:
\begin{eqnarray}
\dltot x_\mu\equal 2\epsilon\cd x\,x_\mu-\epsilon_\mu x^2,\nn\\
\dltot \cc\equal 2(p-3)\epsilon\cd x\,\cc,\nn\\
\dltot U_m\equal-2\epsilon\cd x\,U_m
-\frac{p-1}{2}\frac{kR_p^4}{U^2}\epsilon\cd\del\,U_m,
\label{finalsct}
\end{eqnarray}
with
\bea
k\equiv\frac{2}{5-p},\quad
R_p^2\equiv\sqrt{d_p\cc NU^{p-3}},\quad
d_p\equiv 2^{7-2p}\pi^{(9-3p)/2}\Gamma\Bigl(\frac{7-p}{2}\Bigr).
\eea

The quantum part $\Dlt U_m$ of SCT has an extra numerical factor
$(p-1)/2$ compared to the isometry of the near-horizon geometry of
the D$p$-brane solution \cite{JKY0}.
Since it is the isometry of the near-horizon geometry that determines
the Born-Infeld action with the background, one may wonder if the SCT
derived in SYM is consistent with the Ward-Takahashi identity.
However, it was pointed out in \cite{JKY0} that, since the derivative
of the coupling constant $\eta_\mu\equiv\del_\mu\cc/\cc$
transforms under SCT as
\bea
\dltot\eta_\mu=2(p-3)\epsilon_\mu+\calO(\eta),
\eea
we have to keep terms linear in $\eta_\mu$ in the calculation of the
1-loop effective action to confirm the validity of the Ward-Takahashi
identity.\footnote{
We consider the lowest non-trivial order in $\eta_\mu$
and hence put $\eta_\mu=0$ after the SCT.}
Let us check it for the D-particle case.
For this purpose, we need the quadratic parts of the action:
\bea
&&\calL_{YY}={1\over 2\cc}
Y_{m,ij}(-\del^2+\eta\del+B_{ij}^2)Y_{m,ji},\quad
\calL_{AA}={1\over 2\cc}
A_{ij}(-\del^2+\eta\del+B_{ij}^2)A_{ji},\nn\\
&&\calL_{YA}={2i\over \cc}V_{m,ij}Y_{m,ij}A_{ji},\quad
\calL_{\bar{C}C}=
-i\bar{C}_{ij}(-\del^2+B_{ij}^2)C_{ji},\label{quadratic}\\
&&\calL_{\theta\theta}=\half\theta_{\alpha,ij}
(-\delta_{\alpha\beta}\del+\gamma^m_{\alpha\beta}B_{m,ij})
\theta_{\beta,ji},\nn
\eea
with $V_{m,ij}\equiv\dot B_{m,ij}-\half\eta B_{m,ij}$.
We can read off the 1-loop effective action for the source-probe
situation from (\ref{quadratic}) as
\begin{eqnarray}
\Goneloop\equal N\Tr\Bigl\{10\ln\Bigl(-\del^2+\eta\del+B^2\Bigr)
+\ln\Bigl(1-4V_m\Deta V_m\Deta\Bigr)\nn\\
&&\quad-2\ln\Bigl(-\del^2+B^2\Bigr)
-4\sum_\pm\ln\left(-\del^2+B^2\pm\dot{B}\right)\Bigr\},
\label{one-loop}
\end{eqnarray}
where $\Tr$ denotes the trace over the functional space of $\tau$,
and $\Deta$ is defined by $\Deta\equiv(-\del^2+\eta\del+B^2)^{-1}
=\Delta-\Delta\eta\del\Delta+\calO(\eta^2)$ with
$\Delta\equiv(-\del^2+B^2)^{-1}$.
In (\ref{one-loop}), the first term is the contribution of
$\calL_{YY}$ and $\calL_{AA}$, the second term is due to the mixing
between $Y_m$ and $A$, the third term is the ghost loop and the last
term is from $\calL_{\theta\theta}$.
Keeping only terms independent of and linear in $\eta$, we have
\bea
\Goneloop\equal N\Tr\Bigl\{8\ln\Bigl(-\del^2+B^2\Bigr)
+\ln\Bigl(1-4\dot B_m\Delta\dot B_m\Delta\Bigr)
-4\sum_\pm\ln\Bigl(-\del^2+B^2\pm\dot{B}\Bigr)\nn\\
&&+10\eta\del\Delta+4\eta\Bigl(B_m\Delta\dot B_m\Delta
+2\del\Delta\dot B_m\Delta\dot B_m\Delta\Bigr)
\Bigl(1-4\dot B_\ell\Delta\dot B_\ell\Delta\Bigr)^{-1}\Big\}.
\eea

Note that the SCT of the terms of the form
$\int\d\tau\eta(\tau)\times\mbox{(total derivative terms)}$
in the effective action vanish if we put $\eta=0$ after the
transformation.
Then, since we have
$\bra{\tau}\del\calO\ket{\tau}=(1/2)\del_\tau\!\bra{\tau}\calO\ket{\tau}$
for $\calO=\Delta$ and
$\Delta\dot B_m\Delta\dot B_m\Delta
(1-4\dot B_\ell\Delta\dot B_\ell\Delta)^{-1}$
due to
$\bra{\tau_1}\Delta\ket{\tau_2}=\bra{\tau_2}\Delta\ket{\tau_1}$,
the terms proportional to $\eta$ and relevant for SCT are
\bea
\Tr\Bigl\{4\Bigl(\eta B_m\Delta\dot B_m\Delta\Bigr)
\Bigl(1-4\dot B_\ell\Delta\dot B_\ell\Delta\Bigr)^{-1}\Big\}
=\Tr\Bigl\{4\eta B_m\Delta\dot B_m\Delta+16\eta
B_m\Delta\dot B_m\Delta\dot B_\ell\Delta\dot B_\ell\Delta\Bigr\},
\label{etalinear}
\eea
where we have kept only the terms with the number of derivatives less
than or equal to four.
Note that we cannot adopt the eikonal approximation and drop the
acceleration terms here \cite{HM}, because by integration by parts,
they can be converted to terms without the accelerations.
A method to evaluate it was given in our previous work \cite{HM}:
we took polynomial forms for the background $B_m(\tau)$,
calculated the 1-loop effective action, and identified the result as a
functional of $B_m(\tau)$.
An equivalent, but more refined method was presented in \cite{O}:
all the terms are rearranged into the forms of $f(\tau)\del^m\Delta^n$,
which are calculated using the proper-time representation.
Here we adopt the more convenient method of \cite{O}.
Using the formulas presented in the appendix, the terms linear in
$\eta$ in $\Goneloop$ are calculated to give
\bea
N\int\!\d\tau\eta\biggl(\frac{1}{4}\frac{\dddot{B}\cd B}{B^5}
-\frac{5}{2}\frac{\ddot B\cd B\dot B\cd B}{B^7}
+\frac{5}{4}\frac{\dot B^2\dot B\cd B}{B^7}
+\frac{35}{8}\frac{(\dot B\cd B)^3}{B^9}
+{\rm total\,\,derivative\,\,terms}\biggr).
\eea
After making further the identification of the total derivative terms,
we find the final expression of $\Goneloop$:
\bea
\Goneloop
\equal N\int\!\d\tau\biggl\{-\frac{15}{16}\frac{(\dot{B}^2)^2}{B^7}
+\eta\biggl(\frac{15}{8}\frac{\dot{B}^2\dot{B}\cd B}{B^7}
+{\rm total\,\,derivative\,\,terms}\biggr)\biggr\}.
\label{Goneloop}
\eea

The problem of the extra factor $(p-1)/2$ in $\Dlt U_m$
(\ref{finalsct}) is now resolved by taking into account the SCT of the
term $\eta(15/8)\dot B^2\dot B\cd B/B^7$ in (\ref{Goneloop}).
Moreover, this $\eta$-dependent term in (\ref{Goneloop})
can be eliminated by making the field redefinition
$B_m\to\widetilde{B}_m$ with
\be
\widetilde B_m=B_m-\cc N\frac{3}{4}\frac{\eta\dot B_m}{B^5},
\label{tildeB}
\ee
in $\Gamma_{\rm tree}=\int\d\tau\dot B_m^2/2\cc$.
Then, the SCT for the new variable $\widetilde B_m$ is that of the
isometry without the extra factor $(p-1)/2$.

\section{A new gauge}
Instead of the usual background gauge (\ref{BGgauge}),
here we propose a bit different gauge function useful for discussing
the conformal symmetry in the SYM. In fact, we shall find that the
notorious numerical factor $(p-1)/2$ does not appear this time.
The gauge function of our new gauge is
\bea
G=-\del_\mu A_\mu+\eta_\mu A_\mu+i[B_m,Y_m]+\half\cc b,
\label{NEWgauge}
\eea
which has the additional term $\eta_\mu A_\mu$ compared to the old one
(\ref{BGgauge}).
As in the previous case,
the SCT symmetry broken by $S_{\rm gf+gh}$ can be restored by adding
to the classical SCT the BRST transformation with a field dependent
parameter $\lambda$. In our new gauge, $\lambda[\barC,A]$ is
\bea
\lambda=-4\int\!\d^{p+1}\!x\tr\Bigl(\barC(x)A(x)\cd\epsilon\Bigr).
\label{lambdanew}
\eea
Note that in the usual background gauge (\ref{BGgauge}) the factor
$(p-1)/2$ in $\Dlt U_m$ (\ref{finalsct}) originates in
(\ref{lambda}). However, it is missing in (\ref{lambdanew}).
This implies that we can derive the SCT of the isometry with the correct
factor in our new gauge, namely, $\Dlt U_m$ in the new gauge is given
by
\be
\dltot U_m=-2\epsilon\cd x\,U_m -\frac{kR_p}{U^2}\epsilon\cd\del\,U_m,
\ee
instead of that in (\ref{finalsct}).

Since now the isometry that determines the Born-Infeld action has been
reproduced with the correct coefficient, we can expect that the
$\eta$-dependent terms such as $\eta(15/8)\dot B^2\dot B\cd B/B^7$ in
(\ref{Goneloop}) are missing from the one-loop effective action for
the D-particle.
Let us explicitly check it in the rest of this section.
The quadratic parts of the D-particle action in the new gauge are
\bea
&&\calL_{YY}={1\over 2\cc}Y_{m,ij}(-\del^2+\eta\del+B_{ij}^2)Y_{m,ji},
\quad
\calL_{AA}={1\over 2\cc}A_{ij}(-\del^2-\eta\del+B_{ij}^2)A_{ji},\nn\\
&&\calL_{YA}={2i\over \cc}\dot B_{m,ij}Y_{m,ij}A_{ji},\quad
\calL_{\bar{C}C}=-i\bar{C}_{ij}(-\del^2+\eta\del+B_{ij}^2)C_{ji},
\label{newgaugequadratic}
\eea
and $\calL_{\theta\theta}$ is the same as (\ref{quadratic}) in the old
gauge.
The one-loop effective action for the source-probe configuration
is given by
\bea
&&\Goneloop=
N\Tr\Bigl\{9\ln\left(-\del^2+\eta\del+B^2\right)
+\ln\left(-\del^2-\eta\del+B^2\right)\nn\\
&&\qquad\qquad
+\ln\Bigl(1-4\dot B_m\bigl(-\del^2+\eta\del+B^2\bigr)^{-1}
             \dot B_m\bigl(-\del^2-\eta\del+B^2\bigr)^{-1}\Bigr)\nn\\
&&\qquad\qquad
-2\ln\left(-\del^2+\eta\del+B^2\right)
-4\sum_{\pm}\ln\Bigl(-\del^2+B^2\pm\dot{B}\Bigr)\Bigr\},
\eea
where the origin of the respective terms are the same as before
(\ref{one-loop}).
The $\eta$-independent term in $\Goneloop$ is the same as in
(\ref{Goneloop}), while the term linear in $\eta$ is seen to be
expressed as
\be
\Tr\Bigl\{\eta\del\calO\Bigr\}
=\int\!\d\tau\,\eta(\tau)\half\del_\tau\!\bra{\tau}\calO\ket{\tau},
\label{etaterm}
\ee
in terms of a symmetric $\calO$ satisfying
$\bra{\tau_1}\calO\ket{\tau_2}=\bra{\tau_2}\calO\ket{\tau_1}$.
Eq.\ (\ref{etaterm}) is SCT invariant by putting $\eta=0$ after the
transformation.

Let us summerize our findings in this section.
In the usually adopted background gauge (\ref{BGgauge}), the quantum
SCT has an extra numerical factor $(p-1)/2$. However, if we
adopt the new gauge (\ref{NEWgauge}), this extra factor disappears and
we can derive directly the conformal symmetry that determines the full
Born-Infeld action. Accordingly, the $\eta$-dependent terms in
$\Goneloop$ is also missing up to total derivative terms in our new
gauge.

The calculation in this section shows that the form of quantum SCT
depends crucially on the choice of gauge.
Therefore one may question what the role of the gauge function is.
We can find a clue in \cite{JKY0}, where it is shown that the SCT with
the extra factor $(p-1)/2$ can be related to the SCT without it by
the field redefinition (\ref{tildeB}).
Therefore it seems that the role of the gauge function is to choose
the definition of the fields in the Born-Infeld action side.
In the following sections, we shall study the relationship between the
gauge choice and the redefinition of the fields.

\section{Gauge shifts and field redefinitions}
Here we give some general arguments on the relations between
the change of the gauge function and the field redefinition.
The following is essentially the reproduction of the old arguments
about the independence of the physical S-matrices in the Yang-Mills
theory on the choice of gauge \cite{Lee}.

Let us consider the partition function,
\bea
Z_G[J]=\int\!\calD\phi\,e^{-S+J\cdot\phi},
\eea
where $\phi$ denotes collectively all the fields in the system
and $J$ is the corresponding source.
The action $S$ consists of the gauge-fixing and the ghost terms as
well as the gauge invariant one;
$S=S_{\rm SYM}+i\dBRST(\barC\cd G)$.
The dots in $J\cd\phi$ and $\barC\cd G$ denote both the integration
over the coordinates and the trace over the group indices.

Under an infinitesimal change of the gauge function,
$G\to G+\Delta G$, we have
\bea
Z_{G+\Delta G}[J]-Z_G[J]\equal\int\!\calD\phi
\,\dBRST(-i\bar{C}\cd\Delta G)e^{-S+J\cdot\phi}\nn\\
\equal\int\!\calD\phi\,(-i\bar{C}\cd\Delta G)
(J\cd\delta_{\rm BRST}\phi)e^{-S+J\cdot\phi},
\eea
where in the last equality we have used the BRST Ward-Takahashi
identity. Therefore, the partition function with the gauge function
$G+\Delta G$ is expressed as
\bea
Z_{G+\Delta G}[J]\equal\int\calD\phi
\Bigl(1+(-i\bar{C}\cd\Delta G)(J\cd\delta_{\rm BRST}\phi)\Bigr)
e^{-S+J\cdot\phi}\nn\\
\equal\int\calD\phi
\exp\left\{-S+J\cd\left(\phi+i\delta_{\rm BRST}\phi\,
(\bar{C}\cd\Delta G)\right)\right\}.
\label{ZG+DG}
\eea
This is nothing but the partition function for the field
$\phi+i\delta_{\rm BRST}\phi\,(\bar{C}\cd\Delta G)$
in the original gauge $G$.

Our next task is to restate the property (\ref{ZG+DG}) in terms of the
effective action $\Gamma_G[\varphi]$ which is related to $\ln Z_G[J]$
by the Legendre transformation.
In appendix B, we show that (\ref{ZG+DG}) implies the following
relation between $\Gamma_{G+\Delta G}$ and $\Gamma_G$:
\bea
\Gamma_{G+\Delta G}[\varphi]
=\Gamma_G[\varphi+\langle i\delta_{\rm BRST}\phi\,
(\bar{C}\cd\Delta G)\rangle_\varphi],
\label{gaugeshift}
\eea
where $\langle\calO\rangle_\varphi$ denotes the generating functional
of the 1PI Green's function with an insertion of the operator $\calO$
in the original gauge.

Let us apply the general arguments given above to the relation between
our new gauge (\ref{NEWgauge}) and the old one (\ref{BGgauge}) and
reproduce the field redefinition (\ref{tildeB}) in the D-particle case.
Using (\ref{gaugeshift}) with $\varphi=B_m$ and $\Delta G=\eta A$,
and (\ref{freeprop}), we find that
\bea
\widetilde B_m(\tau)\equal
B_m(\tau)+\Bigl\langle[X_m,C]_{N+1,N+1}(\tau)
\int\!\d\tau_0\tr\bar{C}\eta A(\tau_0)\Bigr\rangle\nn\\
\equal B_m-\int\!\d\tau_0\,\eta(\tau_0)
\Bigl(\Bigl\langle C_{N+1,i}(\tau)\barC_{i,N+1}(\tau_0)\Bigr\rangle
\Bigl\langle Y_{m,i,N+1}(\tau)A_{N+1,i}(\tau_0)\Bigr\rangle
-(i\leftrightarrow N+1)\Bigr)\nn\\
\equal B_m-4\cc N\bra{\tau}\Delta\dot B_m\Delta\eta\Delta\ket{\tau}
=B_m-\cc N\frac{3}{4}\frac{\eta\dot B_m}{B^5},
\label{tildeBagain}
\eea
where we have kept only those terms with the number of derivatives less
than or equal to two.
Note that $\Delta G=\eta A$ can be regarded as infinitesimal since
we are interested only in the lowest non-trivial order in $\eta$.
The result (\ref{tildeBagain}) agrees exactly with (\ref{tildeB})
proposed by \cite{JKY0}.

\section{Matrix model in the $R_\xi$ gauge}
In this section, we shall consider the Matrix model in another gauge.
This will give a support for the validity of the general arguments of
the previous section through a non-trivial calculation.
It will also give a lesson about the structures of the field
redefinitions and the non-renormalization theorem.
The gauge we take here is the $R_\xi$ gauge with $|1-\xi|\ll 1$:
\bea
G=-\del A+i\xi\,[B_m,Y_m]+\half\xi\cc b,
\eea
which has the properties that for $\xi=1$ it reduces to
the original background gauge, and that the $Y$-$A$ mixing is given by
$\calL_{YA}$ of (\ref{quadratic}) independently of the value of $\xi$.
For an infinitesimal  $\alpha\equiv1-\xi$, the difference from the
original background gauge is
\be
\Delta G=\alpha\Bigl(-i[B_m,Y_m]-\half\cc b\Bigr)
=\frac{\alpha}{2}\Bigl(-\del A+i[B_m,Y_m]\Bigr).
\ee
{}From the general formula (\ref{gaugeshift}) of the previous section,
we can obtain the field redefinition relating the $R_\xi$ gauge and
the original background gauge ($\xi=1$). The calculation is carried
out by using the free propagators for $\xi=1$,
\bea
\langle C_{ij}(\tau_1)\bar{C}_{ji}(\tau_2)\rangle\equal
i\bra{\tau_1}\Delta_{ij}\ket{\tau_2},\nn\\
\langle Y_{m,ji}(\tau_1)A_{ij}(\tau_2)\rangle\equal
-2i\bra{\tau_1}\Deta_{ij} V_{m,ij}\Deta_{ij}
\Bigl(1-4V_{n,ij}\Deta_{ij} V_{n,ij}\Deta_{ij}\Bigr)^{-1}\cc(\tau)
\ket{\tau_2},\nn\\
\langle Y_{m,ij}(\tau_1)Y_{n,ji}(\tau_2)\rangle\equal\delta_{mn}
\bra{\tau_1}\Deta_{ij}
\Bigl(1-4V_{\ell,ij}\Deta_{ij}V_{\ell,ij}\Deta_{ij}\Bigr)^{-1}
\cc(\tau)\ket{\tau_2},\nn
\eea
and the formulas given in the appendix A, and taking into account the
derivatives of the coupling $\cc$ carefully.
We find
\bea
\widetilde B_m\equal B_m +\alpha\cc N
\Biggl(\frac{1}{4}\frac{B_m}{B^3}
-\frac{1}{16}\frac{\ddot B_m}{B^5}
+\frac{5}{8}\frac{\dot B\cd B\,\dot B_m}{B^7}
-\frac{5}{16}\frac{\ddot B\cd B\,B_m}{B^7}\nn\\
&&\quad
-\frac{5}{16}\frac{\dot B^2B_m}{B^7}
+\frac{35}{32}\frac{(\dot B\cd B)^2B_m}{B^9}
-\frac{5}{16}\frac{\eta\,\dot B_m}{B^5}
-\frac{5}{8}\frac{\eta\,\dot B\cd B\,B_m}{B^7}
+\frac{3}{16}\frac{\dot\eta\,B_m}{B^5}\Biggr).\label{redefinition}
\eea

To confirm the field redefinition (\ref{redefinition}), let us next
consider the 1-loop effective action in the $R_\xi$ gauge:
\bea
\Goneloop\equal N\Tr\biggl\{
\ln\Bigl((-\del^2+\eta\del+B^2)\delta_{mn}-(1-\xi)B_mB_n\Bigr)
+\ln\Bigl(-\del^2+\eta\del+\xi B^2\Bigr)\nn\\
&&+\ln\biggl(1-4V_m\Bigl(\Deta B_mB_n\Deta\Bigr)V_n
\xi\Bigl(-\del^2+\eta\del+\xi B^2\Bigr)^{-1}\biggr)\nn\\
&&-2\ln\Bigl(-\del^2+\xi B^2\Bigr)
-4\sum_{\pm}\ln\Bigl(-\del^2+B^2\pm\dot B\Bigr)\biggr\},
\label{GoneloopRxi}
\eea
where $\Tr$ for the first term implies also the trace operation
with respect to $(m,n)$.
Keeping only those terms proportional to $\alpha$ and at most linear
in $\eta$, and further with number of derivatives less than or
equal to four, we get after tedious but straightforward calculations
the following result for the shift of the effective action in the
$R_\xi$ gauge from that in the $\xi=1$ gauge:
\bea
&&\int\!\d\tau\alpha N
\biggl(\frac{1}{4}\frac{\dot B^2}{B^3}
-\frac{3}{4}\frac{(\dot B\cd B)^2}{B^5}
-\frac{1}{16}\frac{\dddot B\cd\dot B}{B^5}
-\frac{5}{16}\frac{\dddot B\cd B\,\dot B\cd B}{B^7}
-\frac{5}{8}\frac{\ddot B\cd\dot B\,\dot B\cd B}{B^7}\nn\\
&&\quad
+\frac{35}{8}\frac{\ddot B\cd B(\dot B\cd B)^2}{B^9}
+\frac{105}{32}\frac{\dot B^2(\dot B\cd B)^2}{B^9}
-\frac{315}{32}\frac{(\dot B\cd B)^4}{B^{11}}
+\frac{1}{4}\frac{\eta\,\dot B\cd B}{B^3}\nn\\
&&\quad
+\frac{3}{16}\frac{\eta\,\dddot B\cd B}{B^5}
+\frac{7}{16}\frac{\eta\,\ddot B\cd\dot B}{B^5}
-\frac{5}{8}\frac{\eta\,\ddot B\cd B\,\dot B\cd B}{B^7}
+\frac{5}{16}\frac{\eta\,\dot B^2\dot B\cd B}{B^7}
+\frac{35}{32}\frac{\eta\,(\dot B\cd B)^3}{B^9}\biggr).
\label{shiftGamma}
\eea
In obtaining (\ref{shiftGamma}), we used the cyclicity of the trace to
put all the terms coming from the expansion of (\ref{GoneloopRxi})
into the standard forms of
$\eta B_m\Delta\cdots\Delta f(\tau)\Delta$ or
$\eta\del\Delta\cdots\Delta f(\tau)\Delta$,
and applied the formulas in appendix A.

Then, our next task is to determine the redefinition
$B_m\to\widetilde{B}_m$ in such a way that the sum of the kinetic term
$\dot{B}^2/2\cc$ and the shift of $\Goneloop$ (\ref{shiftGamma})
is identified as the kinetic term of the new field $\widetilde{B}_m$.
The condition for the identification is apparently overdetermined.
For example, the four-derivative terms independent of $\eta$ in
(\ref{shiftGamma}) must correspond to the shift of $B_m$ by some
two-derivative terms. There are 5 kinds of such terms in the shift of
$B_m$; $\ddot B_m/B^5$, $\dot B_m\dot B\cd B/B^7$, $\cdots$,
$B_m(\dot B\cd B)^2/B^9$.
On the other hand, there are 11 kinds of four-derivative terms
independent of $\eta$ in the 1-loop effective action;
$\ddddot B\cd B/B^5$, $\dddot B\cd\dot B/B^5$,
$\cdots$, $(\dot B\cd B)^4/B^{11}$.
Besides, there are 5 kinds of total derivative terms as the ambiguity
of the effective action;
$\d/\d\tau[\dddot B\cd B/B^5]$, $\d/\d\tau[\ddot B\cd\dot B/B^5]$,
$\cdots$, $\d/\d\tau[(\dot B\cd B)^3/B^9]$.
Therefore, we have 11 equations with only 5+5 unknowns.
However, there is a solution and it coincides with
(\ref{redefinition}) obtained from the general formula
(\ref{gaugeshift}).

We can also calculate the quantum SCT in the $R_\xi$
gauge to check the consistency of the field redefinition
(\ref{redefinition}).
Note the expression for the quantum SCT in the $R_\xi$ gauge
is not changed form (\ref{extraterm}), and we obtain
\bea
\Dlt B_m=-\cc N\biggl\{
\Bigl(\frac{3}{2}+\alpha\Bigr)\frac{\epsilon\,\dot B_m}{B^5}
+\frac{5}{4}\alpha\frac{\epsilon\,\dot B\cd B\,B_m}{B^7}
\biggr\}.
\label{qSCTRxi}
\eea
One can easily see that the two quantum SCTs, (\ref{qSCTRxi}) and
(\ref{Dlt_B_final}) with $p=0$, are consistently related by the
redefinition (\ref{redefinition}).

Finally in this section we shall give some comments.
First, although it might seem strange that there appear
two-derivative terms (the first two terms) in (\ref{shiftGamma}),
it does not contradict the non-renormalization theorem of \cite{SSP}.
The statement of \cite{SSP} is that if we remove the
$(\dot B\cd B)^2/B^5$ term by a suitable coordinate transformations,
the other term $\dot B^2/B^3$ will automatically disappear.
Actually, the $B_m/B^3$ term in (\ref{redefinition}) removes
the two terms in (\ref{shiftGamma}) simultaneously.

Our second comment is on the meaning of the field redefinitions.
The redefinition (\ref{tildeB}) is simply a field dependent shift of
the world-volume coordinate $\tau$ and hence is interpretable as a
change of reparametrization gauge in the Born-Infeld action.
However, the redefinition (\ref{redefinition}) contains, besides the
terms interpretable as the target space coordinate transformation
(the $B_m/B^3$ term) and the world-volume reparametrization (terms
containing $\dot B_m$), all kinds of terms with a given dimension.

\section{Conclusions and further directions}
In this paper, we proposed a new gauge which is useful for discussing
the conformal symmetry in the Matrix model.
The form of the quantum modified SCT depends crucially on the choice
of gauge in SYM, and our new gauge reproduced exactly the SCT of the
isometry.
We also gave a general argument on the relation between the change of
gauge and the field redefinition in the effective action.
Then, we examined the $R_\xi$ gauge as an example and reconfirmed the
special nature of the background gauge and our new gauge.
Namely, the diagonal elements of the Higgs fields in SYM correspond
directly to the target space coordinates in the Born-Infeld action
in the static gauge only when we take the background gauge or our new
gauge in SYM.

We shall discuss some further directions of our work.
First, the higher loop analysis of the Matrix model in our new gauge
is an interesting subject.
As the isometry of the near-horizon geometry of the D-brane solution
determines the Born-Infeld action and we can derive the isometry
directly from the 1-loop calculation in SYM in our new gauge, we
expect that the quantum modified SCT in SYM is essentially 1-loop exact.
The analysis of the quantum modified SCT in higher loops
rather than that of the effective action would be a cleverer way to
show the full agreement between the Matrix model and supergravity.
Our new gauge would also be useful for analyzing more general
multi-D-particle systems than the simple source-probe configuration of
this paper.

The new gauge we proposed in this paper may be important even
conceptually.
The fact that the signs of the $\eta\del$ in the kinetic terms
(\ref{newgaugequadratic}) are opposite between $Y_m$ (the coordinates
perpendicular to the branes) and $A_\mu$ (the coordinates parallel to
the branes) reminds us of the spacetime uncertainty principle proposed
by \cite{Y}.
This would be a clue for the understanding of the deep meaning of our
new gauge.

\vspace{.6cm}
\noindent
{\Large\bf Acknowledgments}\\[.2cm]
We would like to thank to our colleagues at Kyoto University for
various useful discussions.
The work of H.\ H.\ is supported in part by Grant-in-Aid for Scientific
Research from Ministry of Education, Science and Culture
(\#09640346).
The work of S.\ M.\ is supported in part
by the Japan Society for the Promotion of Science
under the Predoctoral Research Program.

\vspace{1cm}
\noindent
{\Large\bf Appendices}

\appendix
\section{Useful formulas}
Here we present some useful formulas for calculating the
expectation values using the method of \cite{O}.
Making iterative use of the commutation relations,
\bea
\left[\del,f\right]\equal\dot f,\\
\left[\Delta,f\right]\equal\Delta(\ddot f+2\dot f\del)\Delta,\\
\left[\Delta,\del\right]\equal 2\Delta(\dot B\cd B)\Delta,\nn
\eea
with $f$ being an arbitrary function of $\tau$, we get
\bea
\Delta f\equal f\Delta+2\dot f\del\Delta^2
+4\dot f\dot B\cd B\Delta^3
+\ddot f\Delta^2+4\ddot f\del^2\Delta^3,\label{df}\\
\Delta g\Delta f\equal gf\Delta^2+(2\dot gf+4g\dot f)\del\Delta^3
+(4\dot gf+12g\dot f)\dot B\cd B\Delta^4\nn\\&&
+(\ddot gf+2g\ddot f+2\dot g\dot f)\Delta^3
+(4\ddot gf+12g\ddot f+12\dot g\dot f)\del^2\Delta^4,\label{dgdf}\\
\Delta h\Delta g\Delta f\equal
hgf\Delta^3+(2\dot hgf+4h\dot gf+6hg\dot f)\del\Delta^4
+(4\dot hgf+12h\dot gf+24hg\dot f)\dot B\cd B\Delta^5\nn\\&&
+(\ddot hgf+2h\ddot gf+3hg\ddot f
+2\dot h\dot gf+2\dot hg\dot f+4h\dot g\dot f)\Delta^4\nn\\&&
+(4\ddot hgf+12h\ddot gf+24hg\ddot f
+12\dot h\dot gf+16\dot hg\dot f+32h\dot g\dot f)\del^2\Delta^5,
\label{dhdgdf}\\
\del\Delta f\equal f\del\Delta+\dot f\Delta+2\dot f\del^2\Delta^2
+4\dot f\dot B\cd B\del\Delta^3+3\ddot f\del\Delta^2
+4\ddot f\del^3\Delta^3\nn\\&&
+(4\dot f\ddot B\cd B+4\dot f\dot B^2+4\ddot f\dot B\cd B)\Delta^3
+(8\dot f\ddot B\cd B+8\dot f\dot B^2+24\ddot f\dot B\cd B)
\del^2\Delta^4\nn\\&&
+\dddot f\Delta^2+8\dddot f\del^2\Delta^3+8\dddot f\del^4\Delta^4,\\
\del\Delta g\Delta f\equal gf\del\Delta^2+(\dot gf+g\dot f)\Delta^2
+(2\dot gf+4g\dot f)\del^2\Delta^3,\\
\del\Delta h\Delta g\Delta f\equal hgf\del\Delta^3
+(\dot hgf+h\dot gf+hg\dot f)\Delta^3
+(2\dot hgf+4h\dot gf+6hg\dot f)\del^2\Delta^4,
\eea
where we have dropped the higher derivative terms on the right hand
sides.

Then, we need to calculate $\bra{\tau}\del^m\Delta^n\ket{\tau}$.
First we express it in the proper-time representation as
\be
\bra{\tau_1}\Delta^n\ket{\tau_2}=
\frac{1}{[-\del_{\tau_1}^2+B(\tau_1)^2]^n}\delta(\tau_1-\tau_2)
=\frac{1}{\Gamma(n)}\int_0^\infty\!\d\sigma\sigma^{n-1}
e^{-\sigma[-\del_{\tau_1}^2+B(\tau_1)^2]}\delta(\tau_1-\tau_2).
\ee
This can be evaluated using
\bea
&&e^{-\sigma[-\del_{\tau}^2+B(\tau)^2]}
=\left[1-\sigma^2(\ddot B\cd B+\dot B^2)
-2\sigma^2(\dot B\cd B)\del_\tau
-\frac{8}{3}\sigma^3(\dot B\cd B)^2\right.\nn\\
&&\quad\left.-\frac{4}{3}\sigma^3(\ddot B\cd B+\dot B^2)\del_\tau^2
+2\sigma^4(\dot B\cd B)^2\del_\tau^2
+ \mbox{higher derivative terms}\right]
e^{-\sigma B(\tau)^2}e^{\sigma\del_{\tau}^2},
\eea
obtained from  the Baker-Campbell-Hausdorff's formula, and
\bea
e^{\sigma\del_{\tau_1}^2}\delta(\tau_1-\tau_2)
=\frac{1}{\sqrt{4\pi\sigma}}
\exp\left[-\frac{1}{4\sigma}(\tau_1-\tau_2)^2\right].
\eea
Hence, we get the following derivative expansions for
$\bra{\tau}\Delta^n\ket{\tau}$:
\bea
\bra{\tau}\Delta^2\ket{\tau}\equal\frac{1}{4}\frac{1}{B^3}
-\frac{5}{16}\frac{\ddot B\cd B}{B^7}
-\frac{5}{16}\frac{\dot B^2}{B^7}
+\frac{35}{32}\frac{(\dot B\cd B)^2}{B^9},\nn\\
\bra{\tau}\Delta^3\ket{\tau}\equal\frac{3}{16}\frac{1}{B^5}
-\frac{35}{64}\frac{\ddot B\cd B}{B^9}
-\frac{35}{64}\frac{\dot B^2}{B^9}
+\frac{315}{128}\frac{(\dot B\cd B)^2}{B^{11}},\label{Delta^m}\\
\bra{\tau}\Delta^4\ket{\tau}\equal\frac{5}{32}\frac{1}{B^7},\qquad
\bra{\tau}\Delta^5\ket{\tau}=\frac{35}{256}\frac{1}{B^9},\qquad
\bra{\tau}\Delta^6\ket{\tau}=\frac{63}{512}\frac{1}{B^{11}}.\nn
\eea
As for $\bra{\tau}\del^m\Delta^n\ket{\tau}$, we have,
using $\del^2=-\Delta^{-1}+B^2$ and
$\bra{\tau}\del\Delta^n\ket{\tau}
=\half\del_\tau\bra{\tau}\Delta^n\ket{\tau}$,
\bea
\bra{\tau}\del\Delta^2\ket{\tau}\equal
-\frac{3}{8}\frac{\dot B\cd B}{B^5}
-\frac{5}{32}\frac{\dddot B\cd B}{B^7}
-\frac{15}{32}\frac{\ddot B\cd\dot B}{B^7}
+\frac{35}{16}\frac{\ddot B\cd B\,\dot B\cd B}{B^9}
+\frac{35}{16}\frac{\dot B^2\dot B\cd B}{B^9}
-\frac{315}{64}\frac{(\dot B\cd B)^3}{B^{11}},\nn\\
\bra{\tau}\del\Delta^3\ket{\tau}\equal
-\frac{15}{32}\frac{\dot B\cd B}{B^7},\qquad
\bra{\tau}\del\Delta^4\ket{\tau}=
-\frac{35}{64}\frac{\dot B\cd B}{B^9},\nn\\
\bra{\tau}\del^2\Delta^3\ket{\tau}\equal-\frac{1}{16}\frac{1}{B^3}
-\frac{15}{64}\frac{\ddot B\cd B}{B^7}
-\frac{15}{64}\frac{\dot B^2}{B^7}
+\frac{175}{128}\frac{(\dot B\cd B)^2}{B^9},\label{del^nDelta^m}\\
\bra{\tau}\del^2\Delta^4\ket{\tau}\equal
-\frac{1}{32}\frac{1}{B^5},\qquad
\bra{\tau}\del^2\Delta^5\ket{\tau}=-\frac{5}{256}\frac{1}{B^7},\nn\\
\bra{\tau}\del^3\Delta^4\ket{\tau}\equal
\frac{15}{64}\frac{\dot B\cd B}{B^7},\qquad
\bra{\tau}\del^4\Delta^4\ket{\tau}=\frac{1}{32}\frac{1}{B^3},\qquad
\bra{\tau}\del^4\Delta^5\ket{\tau}=\frac{3}{256}\frac{1}{B^5}.\nn
\eea

\section{Derivation of eq.\ (\ref{gaugeshift})}
In this appendix we present the derivation of eq.\ (\ref{gaugeshift})
for the effective action from eq.\ (\ref{ZG+DG}) for the partition
function $Z_G[J]$. For this purpose, let us introduce the generating
functional of the connected Green's function $W[J,K]$ in the gauge $G$
with sources for $\Delta\phi\equiv i\dBRST\phi\,(\barC\cd\Delta G)$ as
well as for $\phi$:
\be
e^{W[J,K]}=\int\calD\phi\, e^{-S+J\cdot\phi+K\cdot\Delta\phi}.
\ee
Then,
$W_{G+\Delta G}[\widetilde J]\equiv \ln Z_{G+\Delta G}[\widetilde J]$
is expressed in terms of $W[J,K]$ as
\be
W_{G+\Delta G}[\widetilde J]=W[\widetilde J,\widetilde J].
\label{WG+DG}
\ee
Hence, $\Gamma_{G+\Delta G}[\varphi]$, which is the Legendre
transformation of $W_{G+\Delta G}[\widetilde J]$, is expressed using
(\ref{WG+DG}) and the Taylor expansion
$W[J,K]=W[J,0]+ K\cdot(\delta W[J,K]/\delta K)_{K=0}$,
as
\bea
\Gamma_{G+\Delta G}[\varphi]
=\widetilde J\cd\varphi-W[\widetilde J,\widetilde J]
=\widetilde J\cd\varphi-W[\widetilde J,0]
-\widetilde J\cd\frac{\delta}{\delta K}W[\widetilde J,K]\biggr|_{K=0},
\label{GG+DG}
\eea
where the relation between $\varphi$ and $\widetilde{J}$ is given by
\be
\varphi=
\frac{\delta}{\delta \widetilde J}W[\widetilde J,\widetilde J].
\label{phitildeJ}
\ee

On the other hand, let us define $\Gamma[\varphi,K]$ which is the
Legendre transformation of $W[J,K]$ with respect only to $J$:
\be
\Gamma[\varphi,K]=J\cd\varphi-W[J,K],\label{Legendre}
\ee
where $\varphi$ and $J$ are related by
\be
\varphi\equiv\frac{\delta}{\delta J}W[J,K].
\label{phiJ}
\ee
The precise meaning of
$\langle i\dBRST\phi\,(\barC\cd\Delta G)\rangle_\varphi
=\langle\Delta\phi\rangle_\varphi$ in (\ref{gaugeshift}) is in fact
$(\delta\Gamma[\varphi,K]/\delta K)_{K=0}$.
Therefore, the right hand side of (\ref{gaugeshift}) is Taylor
expanded as
\bea
\Gamma_G[\varphi+\langle\Delta\phi\rangle_\varphi]=
\Gamma_G[\varphi]
+\frac{\delta}{\delta\varphi}\Gamma[\varphi,0]\cd
\frac{\delta}{\delta K}\Gamma[\varphi,K]\biggr|_{K=0}.
\label{righthand}
\eea
where we have used $\Gamma_G[\varphi]=\Gamma[\varphi,0]$.

For a given $\varphi$, let $J$ be defined by (\ref{phiJ}) with $K=0$,
and hence $J=\delta\Gamma[\varphi,0]/\delta\varphi$.
In view of the definition (\ref{phitildeJ}) for $\widetilde{J}$,
the difference between $\widetilde J$ and $J$ is
infinitesimal of order $\Delta G$.
Plugging $\widetilde J=J+\Delta J$ into (\ref{GG+DG}) and Taylor
expanding with respect to infinitesimal $\Delta J$, we get
\be
\Gamma_{G+\Delta G}[\varphi]=
J\cd\varphi-W[J,0]-J\cd \frac{\delta}{\delta K}W[J,K]\biggr|_{K=0}
+\Delta J\cd\biggl(\varphi-\frac{\delta}{\delta J}W[J,0]\biggr).
\ee
This agrees with (\ref{righthand}) since we have
$\delta W[J,K]/\delta K=-\delta\Gamma[\varphi,K]/\delta K$
obtained from the derivative of (\ref{Legendre}).
Note that the explicit expression of $\Delta J$ is unnecessary here.

%%%%%%%%%% References %%%%%%%%%%%%%%%%%%%%%%%%%
\newcommand{\J}[4]{{\sl #1} {\bf #2} (#3) #4}
\newcommand{\andJ}[3]{{\bf #1} (#2) #3}
\newcommand{\AP}{Ann.\ Phys.\ (N.Y.)}
\newcommand{\MPL}{Mod.\ Phys.\ Lett.}
\newcommand{\NP}{Nucl.\ Phys.}
\newcommand{\PL}{Phys.\ Lett.}
\newcommand{\PR}{Phys.\ Rev.}
\newcommand{\PRL}{Phys.\ Rev.\ Lett.}
\newcommand{\ATMP}{Adv.\ Theor.\ Math.\ Phys.}
%%%%%%%%%%%%%%%%%%%%%%%%%%%%%%%%%%%%%%%%%%%%%%%%


\begin{thebibliography}{99}
\bibitem{Witten}
E.\ Witten, \J{\NP}{B460}{1996}{335}, hep-th/9510135.
\bibitem{BFSS}
T.\ Banks, W.\ Fischler, S.\ H.\ Shenker and L.\ Susskind,
\J{\PR}{D55}{1997}{5112}, hep-th/9610043.
\bibitem{Susskind}
L.\ Susskind, ``Another Conjecture about M(atrix) Theory'',
hep-th/9704080;\\
A.\ Sen, \J{\ATMP}{2}{1998}{51}, hep-th/9709220;\\
N.\ Seiberg, \J{\PRL}{79}{1997}{3577}, hep-th/9710009.
\bibitem{DKPS}
M.\ R.\ Douglas, D.\ Kabat, P.\ Pouliot and S.\ H.\ Shenker,
\J{\NP}{B485}{1997}{85}, hep-th/9608024.
\bibitem{BBPT}
K.\ Becker, M.\ Becker, J.\ Polchinski and A.\ Tseytlin,
\J{\PR}{D56}{1997}{3174}, hep-th/9706072.
\bibitem{OY}
Y.\ Okawa and T.\ Yoneya, \J{\NP}{B538}{1999}{67}, hep-th/9806108;
\J{\NP}{B541}{1999}{163}, hep-th/9808188.
\bibitem{DEG}
M.\ Dine, R.\ Echols and J.\ P.\ Gray, ``Tree Level Supergravity and
the Matrix Model'', hep-th/9810021.
\bibitem{SSP}
S.\ Paban, S.\ Sethi and M.\ Stern, \J{\NP}{B534}{1998}{137},
hep-th/9805018.
\bibitem{SC}
S.\ Paban, S.\ Sethi and M.\ Stern, \J{JHEP}{9806}{1998}{012},
hep-th/9806028;\\
D.\ A.\ Lowe, \J{JHEP}{9811}{1998}{009}, hep-th/9810075;\\
D.\ A.\ Lowe and R.\ von Unge, \J{JHEP}{9811}{1998}{014},
hep-th/9811017;\\
S.\ Hyun, Y.\ Kiem and H.\ Shin,
``Supersymmetric completion of supersymmetric quantum mechanics'',
hep-th/9903022;\\
S.\ Sethi and M.\ Stern,
``Supersymmetry and the Yang-Mills Effective Action at Finite $N$'',
hep-th/9903049.
\bibitem{O}
Y.\ Okawa, ``Higher-derivative terms in one-loop effective action for
general trajectories of D-particles in Matrix theory'',
hep-th/9903025.
\bibitem{Malda}
J.\ M.\ Maldacena, \J{\ATMP}{2}{1998}{231}, hep-th/9711200.
\bibitem{JKY3}
A.\ Jevicki, Y.\ Kazama and T.\ Yoneya,
\J{\PRL}{81}{1998}{5072}, hep-th/9808039.
\bibitem{JY}
A.\ Jevicki and T.\ Yoneya, \J{\NP}{B535}{1998}{335},
hep-th/9805069.
\bibitem{JKY0}
A.\ Jevicki, Y.\ Kazama and T.\ Yoneya, \J{\PR}{D59}{1999}{06001},
hep-th/9810146.
\bibitem{HM}
H.\ Hata and S.\ Moriyama, ``Conformal Symmetry in Matrix Model beyond
the Eikonal Approximation'', hep-th/9901034, to appear in {\sl \PL}
\bibitem{Lee}
B.\ W.\ Lee, ``Gauge Theories'' in 1975 Les Houches lecture note
``Methods in Field Theory'', R.\ Balian and J.\ Zinn-Justin ed.,
North-Holland (1981).
\bibitem{Y}
T.\ Yoneya, \J{\MPL}{A4}{1989}{1587}.
\end{thebibliography}
\end{document}